\documentstyle[12pt]{article}

\begin{document}
\thispagestyle{empty}
\begin{center}
\Large \tt \bf{Detecting torsion from LISA and teleparallel gravity}
\end{center}
\vspace{2.5cm}
\begin{center} {\large L.C. Garcia de Andrade \footnote{Departamento de
F\'{\i}sica Te\'{o}rica - Instituto de F\'{\i}sica - UERJ
Rua S\~{a}o Fco. Xavier 524, Rio de Janeiro, RJ
Maracan\~{a}, CEP:20550-003 , Brasil.
E-mail : garcia@dft.if.uerj.br}}
\end{center}
\vspace{2.0cm}
\begin{abstract}
Torsion detection from totally skew symmetric torsion waves in the context of teleparallel gravity is discussed. A gedanken experiment to detect Cartan's contortion based on a circle of particles not necessarily spinning is proposed. It is shown that by making use of previous value of contortion at the surface of the Earth computed by Nitsch of $10^{-24} s^{-1}$ a relative displacement of $10^{-21}$ is obtained which is of the order of the gravitational wave of $10^{-3}Hz$. Since LISA has been designed to work in the mHz regime this GW detector could be used for an indirect detection of torsion in $T_{4}$.
\end{abstract}
\vspace{1.0cm}
\newpage

\pagestyle{myheadings}
\markright{\underline{torsion waves and contortion detection}}
\paragraph*{}

The gravitational pp waves in Riemann-Cartan spacetime has been investigated by Adamowicz \cite{1} and by Goenner \cite{2} by making use of Killing vectors. Also Nieto and Ryan \cite{3} have been discussed the motion of spinning particles on the front of a gravitational wave. Later on Hojman et al \cite{4} extended this idea to the non-Riemannian spacetime with propagating torsion. In this paper we show that by considering these ideas in the realm of teleparallelism \cite{5} one is able to obtain a much simpler model which is more akin to physical applications without resource of Killing vectors and any other special sort of symmetries. Let us now turn to the physical analysis of the motion of nonspinning test particles on the plane front of a torsion wave. Thus let us consider the metric perturbation of torsion wave geometry as  perturbation of flat spacetime metric ${\eta}_{ij}$ 
\begin{equation}
g_{ij}={\eta}_{ij}+ h_{ij} 
\label{1}
\end{equation}
where $h_{ij}$ is a small perturbation of the flat Minkowski metric ${\eta}_{ij}$ or $|h_{ij}|<<1$. To simplify matters we shall assume that only $A_{X}$ mode survives while amplitudes of the type $A_{+}$ vanish.  Since the metric $g_{ij}$ changes in time with the passage of the wave this causes the proper distance between the spinning test particles \cite{3} will change as well with the passage with torsion wave. For a plane wave propagating along the $z-axis$, the proper distance in the $xy$-plane \cite{6} is given by
\begin{equation}
dl = [(1+{h^{TT}}_{xx})dx^{2}+ (1-{h^{TT}}_{xx})dy^{2}+2 {h^{TT}}_{xy}dx dy]^{\frac{1}{2}}
\label{2}
\end{equation}
where the metric perturbation components are
\begin{equation}
{h^{TT}}_{xx}= -{h^{TT}}_{yy}=A_{+} e^{-i{\omega}(t-z)}
\label{3}
\end{equation}
\begin{equation}
{h^{TT}}_{xy}= A_{X} e^{-i{\omega}(t-z)}
\label{4}
\end{equation}
The Riemann curvature components are obtained by making use of the $T_{4}$ condition on vanishing of the RC curvature tensor. This yields 
\begin{equation}
R_{1010}=[{\partial}_{1}K_{010} - {\partial}_{0}K_{110}]
\label{5}
\end{equation}
\begin{equation}
R_{1310}=[{\partial}_{3}K_{110} - {\partial}_{1}K_{310}]= -\frac{1}{2}\ddot{h_{+}}= -\frac{{\omega}^{2}}{2}A_{+}e^{-i{\omega}(t-z)}
\label{6}
\end{equation}
\begin{equation}
R_{1313}=[{\partial}_{3}K_{311} - {\partial}_{1}K_{313}] 
\label{7}
\end{equation}
\begin{equation}
R_{2020}=[{\partial}_{0}K_{220} - {\partial}_{2}K_{020}]
\label{8}
\end{equation}
\begin{equation}
R_{2320}=[{\partial}_{3}K_{220} - {\partial}_{2}K_{320}]
\label{9}
\end{equation}
\begin{equation}
R_{2323}=[{\partial}_{3}K_{223} - {\partial}_{2}K_{323}]
\label{10}
\end{equation}
\begin{equation}
R_{1020}=[{\partial}_{0}K_{120} - {\partial}_{1}K_{020}]= -\frac{1}{2}{\ddot{h}}_{X}
\label{11}
\end{equation}
\begin{equation}
R_{1320}=[{\partial}_{3}K_{120} - {\partial}_{1}K_{320}] 
\label{12}
\end{equation}
\begin{equation}
R_{1023}=[{\partial}_{3}K_{120} - {\partial}_{1}K_{023}]= -\frac{1}{2}{\ddot{h}}_{X}
\label{13}
\end{equation}
\begin{equation}
R_{1323}=[{\partial}_{3}K_{123} - {\partial}_{1}K_{323}]= -\frac{1}{2}{\ddot{h}}_{X}
\label{14}
\end{equation}
Due to the totally skew-symmetry of the contortion tensor in the neutrino wave components such as $K_{121}$,$K_{223}$,$K_{220}$,$K_{311}$ vanish which allows us to simplify the Riemann curvature expressions in order to obtain equations of the type
\begin{equation}
{\partial}_{0}K_{123}= -\frac{1}{2}{\ddot{h}}_{X}
\label{15}
\end{equation}
which yields the following complex solution
\begin{equation}
K_{123}= -\frac{i{\omega}}{2}A_{X}e^{-i{\omega}(t-z)}
\label{16}
\end{equation}
when one takes the real part of the complex phase this expression reduces to
\begin{equation}
Im(K_{123})=- \frac{A_{X}}{2}{\omega}cos({\omega}(t-z))
\label{17}
\end{equation}
where the Im denotes the imaginary part of the complex representation of the torsion wave. When the torsion wave passes by a ring of spinning particles the perturbation of the ring in the $xy$-plane is
\begin{equation}
{\delta}l^{X} = -\frac{A_{X}}{2}l cos({\omega}(t-z))
\label{18}
\end{equation}
Thus contortion contributes to ${\delta}l$ which is the change of the separation of the spinning particles when the torsion wave hits the ring.
Actually is better to consider the relative displacement of particles in the ring of particles which is given by $\frac{{\delta}l}{l}$ which by the expressions (\ref{16}) and (\ref{17}) yields
\begin{equation}
|Im(K_{123})|= {\omega}\frac{{\delta}l^{X}}{l}
\label{19}
\end{equation}
Note that by considering a frequency of the order of ${\omega}=10^{-3}Hz$ and the value obtained by Nitsch \cite{7} for the contortion at the surface of the Earth of $K_{123}=10^{-24} s^{-1}$ one obtains from expression (\ref{18}) a relative displacement for the ring of 
\begin{equation}
\frac{{\delta}l^{X}}{l}=10^{-21}
\label{20}
\end{equation}
which is of the order of the gravitational wave Peres gedanken experiment result. Let us now consider $T_{4}$ amplitude of perturbations of spacetime in terms of contortion in general case to express the analogous of the gravitational field energy momentum tensor. This model yields the following Riemann curvature tensor w.r.t contortion
\begin{equation}
R^{a}_{{0}{b}{0}}(g)={\partial}^{a}{K_{0b 0}}-{\partial}_{0}{{K}^{a}}_{b 0}
\label{21}
\end{equation}
where here g is the symbolic representation of the metric tensor $g_{ij}$. By considering that the contortion is just a function of time and that the metric curvature can be expressed as 
\begin{equation}
R_{a0b0}(g)=-\frac{1}{2}{{\ddot{h}}^{TT}}_{ab}= -{\partial}_{0}{K_{ab0}}
\label{22}
\end{equation}
which allows us to write the contortion in terms of the acceleration of metric perturbation h as
\begin{equation}
\frac{1}{2}{{\dot{h}}^{TT}}_{ab}= {K_{ab0}}
\label{23}
\end{equation}
which allows us to find the tensor or GW perturbation in terms of the Cartan contortion tensor as
\begin{equation}
{{h}^{TT}}_{ab}= 2\int{{K_{ab0}}dt}
\label{24}
\end{equation}
Substitution of this last expression into the expression for the pseudo-tensor for the GW field 
\begin{equation}
{t^{GW}}_{ij}=\frac{1}{32{\pi}}<{{h}^{TT}}_{ab,i}{{{h}^{TT}}^{ab}}_{,{j}}>
\label{25}
\end{equation}
yields
\begin{equation}
{t^{GW}}_{{0}{0}}=\frac{1}{32{\pi}}<{{h}^{TT}}_{ab,0}{{h}^{TT}}^{ab}_{,{0}}>
\label{26}
\end{equation}
or
\begin{equation}
{t^{GW}}_{00}=\frac{1}{8{\pi}}< K^{2}>
\label{27}
\end{equation}
where $K^{2}= K_{ab0}K^{ab0}$ where $a,b=1,2,3$. This expression could be also obtained directly from the expression derived by de Andrade et al. \cite{8} for the energy-density of the gravitational field
\begin{equation}
{t}_{{i}{j}}=h^{A}_{i} j_{A{j}}+\frac{c^{4}}{4{\pi}}{{\Gamma}^{l}}_{{i}{k}} {{S_{l}}^{k}}_{j}
\label{28}
\end{equation}
in the absence of gauge current $j^{l}_{A}$ where tetrad indices are represented by latin capital letters. Actually is is not difficult to show that contortion contribution to the gravitational field energy (\ref{26}) is contained into expression (\ref{27}). This can be obtained as follows. Consider that the $00-component$ of (\ref{27}) can be expressed as
\begin{equation}
{t}_{{0}{0}}=h^{A}_{0} j_{A{0}}+\frac{c^{4}}{4{\pi}}{{\Gamma}^{l}}_{{0}{k}}{{S_{l}}^{k}}_{0}
\label{29}
\end{equation}
and since 
\begin{equation}
S^{lk0} = \frac{1}{2}[K^{lk0}-g^{l0}{{T^{p}}_{p}}^{k} +g^{k0}{T^{pl}}_{p}] 
\label{30}
\end{equation}
where $T^{ijk}$ is the torsion tensor we notice that the second term of expression (\ref{27}) reduces to the contortion squared expression. From expressions above and the teleparallel condition one is now able to conclude that the transverse perturbation in the case of the gravitational wave (GW) is constraint also by the contortion K and its derivatives. Let us now compute two examples of the amplitude of the metric perturbations in terms of contortion component. In the first case we compute the amplitude for the Earth-Sun system where torsion is computed \cite{3} to be $K=10^{-19} s^{-1}$ for a frequency of ${\omega}_{0}=10^{-3} Hz$. With this aim we make use of the expression (\ref{25}) in the form
\begin{equation}
K = \frac{h}{2}{\omega}_{0}
\label{31}
\end{equation}
This formula is exactly the same obtained for the torsion wave above. Substitution of the above data into this expression yields an perturbation amplitude of $h= 2.10^{-16}$. Despite of this very nice value it is very difficult to perform such a experiment since since at this low frequency the planetary motion introduces high perturbations in spacetime which destroy any small perturbations in spacetime \cite{6}. From the experimental known data ${\omega}_{0}=10^{4} s^{-1}$ and making use of the Nitsch value for the Earth contortion $K=10^{-24}s^{-1}$ \cite{3} one is able to compute another example based on the formula (\ref{30}) which yields this time an amplitude perturbation of the order $h= 10^{-28}$ which is beyond the best values obtained from GW detectors so which is of the order of $10^{-22}$. Yet some torsion value at the surface of the Earth was computed by Rumpf \cite{9} using axial symmetric sources like $S=10^{-15} s^{-1}$. With this new value for the torsion at the surface of the Earth one is able to compute an amplitude of the order $h=10^{-19}$ for a frequency of $10^{4} Hz$. This value is very well within some known detectors. A more detailed discussion of the possibility of designing these kind of experiments as well as a detector of spin polarized to test torsion theories of gravity where torsion propagates were previously discussed by Hojmann, Rosembaum and Ryan (HRR) \cite{4}. Actually HRR have suggested that it is possible to find an energy flow required to excite a cold detector in nearly steady-state vibrations possesing energies at room temperature thermal vibrations. Taking into account a detector of Weber type , for a frequency of the order $10^{3} Hz$ torsion would have to be of the order of $10^{21} cm^{-1}$ which is an unrealistic value that has never been observed. Thus HRR have reached the conclusion that GW experiments that have not detected excitations of this magnitude, actually gives us the information that this huge number could be taken as a rough upper limit for torsion. Since this number seems to be extremely high even for very massive astrophysical sources such as black holes and binary pulsars, our model seems to be much more realistic which again shows that teleparallel gravity even today maybe useful not only to explain important physical problems such as the issue of the gravitational field energy but also help us to understand some pratical problems such as torsion and gravitational waves. Other interesting application is the generalization of Nieto and Ryan work \cite{3} of the motion of spinning particles on a plane GW to $T_{4}$. Other interesting terrestrial proposals to torsion detection have been recently appeared in the literature by Lammerzahl \cite{10} and myself \cite{11}.

\newpage

\section*{Acknowledgements}
 I would like to express my gratitude to Prof. P. S. Letelier, Prof.O.Aguiar,Prof. J.G.Pereira,  Prof. M. Rosenbaum and Prof.M.P.Ryan for helpful discussions on the subject of this paper. Financial support from CNPq. is grateful acknowledged.

\end{document}